# Magnetic frustration and paramagnetic state transport anomalies in Ho$_4$RhAl and Er$_4$RhAl: Possible test cases for newly identified roles of itinerant electrons

Ram Kumar,[1,*] and E.V. Sampathkumaran[2,3,**]

[1]*Tata Institute of Fundamental Research, Homi Bhabha Road, Colaba, Mumbai 400005, India*
[2]*Homi Bhabha Centre for Science Education, Tata Institute of Fundamental Research, V. N. Purav Marg, Mankhurd, Mumbai, 400088 India*
[3]*UGC-DAE-Consortium for Scientific Research, Mumbai Centre, BARC Campus, Trombay, Mumbai 400085, India*

Abstract

We report the results of magnetic, heat-capacity, electrical and magnetoresistance measurements on Ho$_4$RhAl and Er$_4$RhAl, characterized by 3 sites for rare-earths (R). The main conclusions are: (i) Antiferromagnetic ordering sets in at ($T_N$=) ~ 8.8 and ~ 4.0 K respectively. While Ho compound appears to enter into a complex spin-glass phase at $T < T_N$ (at nearly 5 K), spin-glass component appears to set in essentially almost at $T_N$ for the Er case. (ii) The loss of the spin-disorder contribution in the magnetically ordered state is not pronounced, mimicking that in Gd$_2$PdSi$_3$, a compound which now attracts interest in the area of topological Hall effect and magnetic skyrmions, indicating complex Fermi surface. (iii) There is a minimum in the temperature dependence of electrical resistivity in the case of only Ho above $T_N$, but significant negative magnetoresistance is observed over a wide temperature range in the paramagnetic state increasing with decreasing temperature for both the cases. This finding establishes that these compounds belong to a select group of intermetallics in which spin-disorder contribution apparently increases gradually as one approaches respective $T_N$ with decreasing temperature. This could be an experimental signature for the effect due to classical spin-liquid above $T_N$. In view of these properties analogous to those of Gd$_2$PdSi$_3$, it is of interest to investigate these 4:1:1 compounds further to understand possible unconventional roles of itinerant electrons, not only in the magnetically ordered state but also in the paramagnetic state, predicted by some theories in recent years for which this Gd compound is considered to be a classic example.

Besides, magnetoresistance and isothermal entropy change (magnetocaloric effect) track each other. Isothermal entropy change at the peak in its plot versus temperature is large in comparison with that of the Gd analogue, possible implication of which is pointed out.

*ramasharamyadav@gmail.com
**Corresponding author: sampathev@gmail.com

## 1. Introduction

For the past several decades, the indirect exchange interaction called 'Ruderman-Kittel-Kasuya-Yosida (RKKY)' interaction in metallic magnetic materials has been known as a mechanism to mediate magnetic interaction among various magnetic sites. Recently, there is a major deeper understanding with respect to this simplistic approach, in the sense that the magnetic interaction mediated by itinerant electrons has been realized to result in unconventional, conceptually interesting (i) magnetic ground states [1] and (ii) electrical transport anomalies [2], in metallic environments.

With respect to (i), the magnetic state due to the RKKY interaction can be frustrated, a kind of situation which has been proposed to lead to magnetic skyrmions – an exotic magnetic textures of great current interest. In simple terms, the ground state can not be satisfied simultaneously by all the magnetic interactions of the local moment with the conduction band. This frustration is different from the traditionally known 'geometrical frustration' concept, well-studied among many insulators, in the sense that the latter is due to certain local arrangement of magnetic ions in the lattice [like the triangles or tetrahedra with inter-site antiferromagnetic (AF) interaction]; therefore such a frustration is short-ranged in nature, unlike the RKKY interaction via the conduction electrons which is long-ranged in nature; naturally the band structure features play a role on frustration. Needless to state that, one sees the signatures of competition between ferromagnetic (F) and AF interaction in the measured properties.

With respect to (ii), the RKKY interaction coupled with a tendency for frustrated magnetism (even geometrically) has been recently predicted [2] to lead to an unexpected upturn in electrical resistivity ($\rho$) in the paramagnetic state before the onset of long-range magnetic order, triggered by previous experimental observations - that is, the Kondo-like transport anomalies in rare-earth (R) intermetallics in which 4f electrons are strictly localized [3, 4]. Thus, the origin of this upturn is different from that in the Kondo alloys. This theory also predicts that this upturn persists even in the dilute limit of magnetic ions (analogous to the single-ion Kondo effect), which finds support in the transport behavior of $Gd_{2-x}Y_xPdSi_3$ [Ref. 3]. Such features are induced by 'classical spin-liquid behavior' [2]. [Incidentally, we cautioned more than two decades ago [3] that such factors - which we labelled "magnetic precursor effect" - not realized at that time, might also contribute to low-temperature upturns in $\rho$ and heat-capacity ($C$) in apparently non-Fermi-liquid systems].

Interestingly enough, the ternary compound $Gd_2PdSi_3$ – the magnetic and transport anomalies (including topological-Hall-effect-like Hall anomalies, Refs. 3, 5, 6), which were pioneered by one of us more than two decades ago – has been proposed to be a classic example for both the novel roles of itinerant electrons. These anomalies led to the studies of angle-resolved photoemission spectroscopy (ARPES) on this compound [7], which provided evidence for a complex nesting of Fermi surface in the magnetically ordered state. Though a report based on first principle calculations [8] claims that the spin structure of this compound does not have its origin in Fermi surface nesting, there is no experimental evidence till now in favour of this prediction. Clearly, in-depth investigations of more metallic compounds, containing such well-localized 4f orbitals (that is, heavy R systems) in which complications due to 4f-delocalization (like the Kondo effect in light rare-earths) do not exist, are warranted for theoretical advancement in magnetism.

In this respect, recently we have reported the results of extensive bulk investigations on some members of the ternary rare-earth families, $R_4PtAl$ (R= Gd, Tb, and Dy) [Refs. 9-11] and $R_4RhAl$ (Gd and Tb) [Ref. 12], forming in $Gd_4RhIn$-type cubic structure (F$\bar{4}$3m). This crystal structure is characterized by 3 sites for R [see, for example, Refs. 13-16]. Incidentally, novelty of coordination features and magnetic characteristics were reported by Doan et al [13] for the Cd based family. In the above-mentioned compounds, we found a variety of exotic magnetic frustration features as well as thermal, transport and magnetocaloric effect (MCE) anomalies, due to the competition between AF and F interactions. In view of this, we considered it important to extend these studies to other members of these ternary families. In this article, we report the results of our studies on isostructural $Ho_4RhAl$ and $Er_4RhAl$, comparing the observed properties with those of others. We make many interesting observations, in particular the ones relevant to the above theoretical ideas. Needless to state that, following initial synthesis reports, there is no detailed work on these RE-Rh-Al based ternary families.

## 2. Experimental details

The polycrystalline specimens of Ho$_4$RhAl and Er$_4$RhAl were synthesized by arc-melting together requisite amounts of high purity constituent elements [R: >99.9%; Rh: >99.99%; Al: >99.999%] in a water-cooled copper hearth in an atmosphere of argon (pressure: ~ 10 torr). In order to attain homogeneity, the ingots were flipped and melted four times. The molten ingots were subsequently annealed in evacuated sealed quartz tubes at 750C for two weeks. Powder x-ray diffraction patterns obtained using Cu K$_\alpha$ radiation are shown in Fig. 1 and the full-width at half-maximum, say for the intense lines in the range (2θ=) 30º to 40º, is typically 0.15º. The results of Rietveld analysis are also shown. These establish that the materials are single phase forming in the Gd$_4$RhIn-type cubic structure. No impurity peak attributable to any other phase could be seen within the detection limit (<2%) of x-ray diffraction method. Scanning electron microscopic technique was also employed to verify composition homogeneity. *D*c magnetic susceptibility ($\chi$) (1.8 – 300 K) as well as isothermal magnetization (*M*) measurements at selected temperatures (*T*) were performed on an ingot of about 10 mg employing a commercial (Quantum Design) SQUID magnetometer; *T*-dependence of ac $\chi$ was obtained with an ac field of 1 Oe with four frequencies (1.3, 13, 133 and 1333 Hz). *T*-dependence of heat-capacity was recorded down to 1.8 K in the vicinity of magnetic ordering temperatures by relaxation method with the help of a commercial Physical Properties Measurements System (PPMS, Quantum Design). Zero-field and in-field dc $\rho$(*T*) behavior (1.8 – 300 K) were performed by the four-probe method with the same PPMS. The data collection was usually done for the zero-field-cooled (ZFC) state of the specimens.

## 3. Results

### 3.1 *Properties of Ho$_4$RhAl*

In Fig. 2, we show the results of dc $\chi$ measurements in the *T*-range 1.8-300 K, obtained in two different magnetic fields (*H*), 100 Oe and 5 kOe. As inferred from Fig. 2a from the data measured in 5 kOe, $\chi$ follows Curie-Weiss behavior over a wide temperature range in the paramagnetic state and the value of the effective magnetic moment ($\mu_{eff}$) obtained from the Curie-Weiss fitting above 25 K is ~10.7 $\mu_B$/Ho, in good agreement with that expected for trivalent Ho ions (10.6 $\mu_{eff}$). This suggests that Rh is non-magnetic. The paramagnetic Curie-temperature ($\theta_p$) is ~-18 K and the negative sign implies the existence of AF correlations among Ho ions. As the temperature is lowered below 20 K, a weak shoulder develops at ~8.8 K (which one can see clearly in the derivative curve, not shown here), possibly indicating the onset of magnetic ordering [For further support, see below]. If so, the ratio of $\theta_p$ to Néel temperature ($T_N$), called "frustration ratio", is nearly 2, which is a signature [17] of the existence of magnetic frustration. A knowledge of crystal-field effects is desirable to delineate the contribution from frustration to the reduced $T_N$ value. We carried out additional dc *M* measurements to establish [18] the role of frustration. The $\chi$(*T*) curve obtained in a field of 100 Oe for the ZFC condition exhibits a distinct peak at ~5 K (which is also seen in the curve for 5 kOe). The point to be noted is that the ZFC and field-cooled (FC) curves tend to deviate below at $T_N$ ~8.8 K, as shown in Fig. 2b. The peak temperature corresponds to freezing temperature. The FC curve, instead of flattening below 5 K as in conventional spin-glasses, keeps increasing below 5 K, which is a characteristic feature of cluster-glass systems [9, 11, 19-21]. We have also performed isothermal remnant magnetization ($M_{IRM}$) studies at selected temperatures (1.8 and 5 K) by measuring the magnetization as a function of time (*t*) after switching of the field (in this case, 5 kOe) for the zero-field state of the specimen. The $M_{IRM}$ curves, thus obtained, exhibit a slow delay with time, as shown Fig. 2c. Since the curves overlap for both the temperatures, we show the same for one temperature only. We could fit the curves to a sum of two exponential functions of the type $M_{IRM}= M_{IRM}(0) + Ae^{-(t/\tau)}$, where A is a constant and τ is the relaxation time, – and not to a single exponential function as known for many spin-glasses – implying the existence of two relaxation times (τ) (~200s, ~2000s); large τ values imply cluster type frozen magnetic state.

To gain further knowledge about the glassy state, we have measured ac χ in the vicinity of magnetic transitions. In the absence an external dc field, the real part (χ') exhibits a broad peak around 6 K, and an additional weak structure (dip) at 4.8 K. A weak shoulder is also visible at $T_N$ for 1.3 Hz (Fig. 3a). Clearly the overlap of the features due to AF and spin-glass gives rise to complex ac χ curves. There is also a weak frequency dependence with the peak-temperature increasing (in this case by about 0.5 K) as the frequency is varied from 1.3 Hz to 1333Hz – characteristic of spin-glasses. These features imply that this compound is characterized by reentrant spin-glass behavior. Different from the well-known spin-glasses, an external field of 5 kOe does not depress the peak completely; this can be attributed to the non-glassy magnetic phase setting in near 8.8 K, leading to a broad χ' peak, continuing to coexist with the glassy phase in zero field down to 1.8 K. Complexity of the glassy phase is obvious from the fact that the imaginary part (χ'') does not exhibit a peak, as the values keep increasing with decreasing temperature, as shown in Fig. 3b.

The onset of long-range antiferromagnetic order at 8.8 K is confirmed by the heat-capacity data shown in Fig. 4a. A well-defined λ-anomaly with the upturn appearing below about 10 K is apparent in the zero-field $C(T)$ curve, with the peak occurring at 8.3 K. No other peak is visible at lower temperatures, say, around 5 K, which is not inconsistent with the glassiness of ordered state inferred above from the magnetization data around this temperature. With the application of $H$, the peak temperature gets depressed, say, to 7.6 K for 30 kOe and the peak is completely suppressed for 50 kOe, however with a significant change in slope of the curve appearing at a lower temperature (6 K). This suppression of the peak temperature by the external field establishes that the onset of magnetic order is of an antiferromagnetic state. We have derived isothermal entropy change [$\Delta S = S(H)-S(0)$] – a measure of magnetocaloric effect – by integrating $\Delta C/T$, the results of which are shown in Fig. 4b for a change of field from zero to 10, 30 and to 50 kOe. The values for 10 kOe are found to be too small to attach any significance. It is to be noted that the sign of -$\Delta S$ is positive, which is a signature [22] of a tendency for spin-reorientation towards ferromagnetic alignment at these fields. Notably, the maximum appears at a temperature slightly higher than $T_N$, remaining positive even at 20 K, as though ferromagnetic clusters behaving like classical spin-liquid start forming well above $T_N$. It is to be remarked that the value at the maximum (~ 2.4 and 6.3 J/kG K, for 30 and 50 kOe) is much higher than that noted for the Gd analogue [12].

We have measured isothermal $M$ up to 70 kOe at 1.8, 3 and 5 K, which provide evidence for spin-reorientation inferred from isothermal $\Delta S$ data above. At 1.8 K, $M$ varies essentially linearly till about 20 kOe beyond which there is an upward curvature in the range 20-30 kOe, in support of spin-reorientation; at further high fields, the slope of the curve remains essentially intact (Fig. 5). We could not trace any hysteresis and therefore there is no evidence for disorder-broadened first-order field-induced transitions, noted for other members in this family [9-12]. The curves for 3 and 5 K almost overlap with that for 1.8 K and hence are not shown. Occurrence of spin-reorientation is consistent with the antiferromagnetism in the virgin state.

The results of electrical resistivity and magnetoresistance (MR) are shown in Fig. 6. As expected for metallic systems, dρ/d$T$ is positive down to about 15 K in the paramagnetic state (Fig. 5a). However, this is followed by an upturn, which is usually not expected for heavy rare-earths-based metallic systems. One of us reported some exceptions in some families, e.g., $Gd_2PdSi_3$ [3], RCuAs$_2$ [Ref. 4], $Gd_2CuGe_3$ [Ref. 23], $R_7Rh_3$ [Ref. 24], $Tb_5Si_3$ [Ref. 25], and $Gd_3RuSn_6$ [26], apart from $R_4PtAl$ [9-11]. This upturn is intercepted by a kink at $T_N$. Thereafter, ρ remains essentially constant (showing a decrease of less than 0.1%) with a lowering of temperature to 1.8 K. Thus, there is no notable fall of ρ due to the loss of spin-disorder contribution at the onset of Néel order in the present compound. This kind of behavior, mimicking that of $Gd_2PdSi_3$ [Ref. 3], has been known [27] to arise from complicated Fermi surface in the magnetically ordered state. We have taken ρ($T$) data in the presence of external fields and this feature persists even at fields as high as 30 kOe and no significant fall at the onset of magnetic order is visible even at higher fields of 50 and 70 kOe. Returning to the paramagnetic state behavior, there is a significant suppression of ρ by $H$, and the magnitude of MR, defined as [ρ($H$)-ρ(0)]/ρ(0), increases with decreasing temperature, as shown in an inset of Fig. 6a for $\Delta H= 50$ kOe, from $T$ much greater than $T_N$. We have also obtained isothermal MR data and the results are shown in Fig. 6b for 1.8, 8, 10, 20, and 50 K. As in the case of isothermal magnetization, the curves are non-hysteretic. The magnitude of MR for $H= 70$ kOe is as high as about 14% at 20 K, far above $T_N$. The sign of MR remains negative not only in the paramagnetic state (as expected), but also in the magnetically ordered state. There is no evidence for positive MR just below $T_N$, even at low fields (which should appear in simple antiferromagnets), and this makes antiferromagnetism of this compound different from other members of this family [9-12]. While spin-glass phase (at 1.8 K) can lead to negative MR sharply varying at low fields [28], the absence of positive MR in the AF phase at 8 K implies a Fermi surface with magnetic gaps. The functional form of MR($H$) below $T_N$ is quite revealing. While MR varies as square of $H$ in the low-field range (<20 kOe) as in the paramagnetic state curves (20 and 50 K), we have noted a deviation from this functional form at higher fields and at lower temperatures below $T_N$, which is attributable to field-induced spin-reorientation. The persistence of quadratic field-dependence in the magnetically ordered state at low fields is interesting and needs to be understood theoretically. Overall, these transport data reveal novel magnetic state of this compound.

*3.2. Properties of Er$_4$RhAl*

The results of measurements of dc $\chi(T)$ as well as of $M_{IRM,}$ and isothermal magnetization at 1.8 K are shown in Fig. 7. There is a distinct peak at 3.5 K, indicating the onset of magnetic ordering of an AF-type. As shown in Fig. 7a, inverse $\chi$ measured in a field of 5 kOe varies linearly over a wide $T$-range above 10 K and the values of $\mu_{eff}$ and $\theta_p$ derived from this Curie-Weiss region turn out to be ~9.9$\mu_B$/Er (in close agreement with the theoretical value of 9.59 $\mu_B$ for trivalent Er ion) and ~ -16 K respectively; marginally higher value can arise from conduction electron polarization by Er moments. Though AF ordering is consistent with the negative sign of $\theta_p$, the ratio $\theta_p/T_N$ (16/3.5) is large, which is consistent with magnetic frustration. As shown in Fig. 7b, $\chi(T)$ curves for ZFC and FC conditions obtained in a low-field of 100 Oe show a significant deviation below the peak, which is a characteristic feature of spin-glasses. Since we are not able to resolve additional magnetic transitions around 3.5 K in these measurements, we wonder whether spin-glass freezing (indicated by ZFC-FC deviation) and AF ordering are coterminous (or occur at marginally different temperatures) in this compound. A careful look at the ZFC and FC data suggests that these two curves indeed weakly deviate below 10 K (that is, above $T_N$) and this is associated with the magnetic state before long-range ordering sets in (see Discussions). $M_{IRM}$ curves (Fig. 7c) measured at 1.8 and 3 K decay with time consistent with spin-glass freezing. As in the case of Ho compound, a sum of two stretched exponential function is required to get a good fit of the curve; that is, there are two relaxation times, the values of which are large (~250, ~2000 s). But we indeed found a weak decay even at 5 K (but not at 10 K), as though the magnetic state building above $T_N$ (inferred from ZFC-FC $\chi$ data) exhibit glassiness. We have also made a comparison of the magnitudes of the change in $M_{IRM}$ after 1h; the net change in the value is about 20 times weaker for 5 K with respect to that for 3 K, and therefore possible glassiness just above $T_N$ is quite weak. We are not able to resolve hysteresis in the isothermal $M$, measured at 1.8, 3 and 5 K (see Fig. 7d for the curve at 1.8 K); $M$ varies rather gradually with $H$ as shown in Fig. 7d. Though we could not clearly resolve a feature due to spin-reorientation in this data, these behaviors of $M(H)$ are sufficient to suggest dominance of AF ordering, and not of ferromagnetic ordering.

Figs. 8a and 8b show ac $\chi$ behavior. In the absence of an external dc magnetic field, there is a peak at about 4 K in $\chi'$. The frequency dependence, though extremely weak, is definitely present. This is more apparent from the peak temperature of $\chi''$ at various temperatures. Though the curves are quite noisy, the frequency dependence is apparent. In the presence of 5 kOe, the peak in $\chi''$ vanishes (as expected for spin-glasses), but the peak is not completely suppressed in $\chi'$; that is, a broad peak, overlapping for all frequencies, persists at a slightly depressed temperature of 3 K. This $\chi'$ feature in 5 kOe therefore is attributed to a non-glassy phase. These findings appear to endorse the view (see above) that AF phase sets in along with glassy phase in the close vicinity of 4 K.

In Fig. 8c, $C(T)$ behaviors in zero field as well as in 30 and 50 kOe are shown below 25 K. The existence of a λ-anomaly with a peak endorses the onset of long range magnetic order around 4.5 K (that is, at the middle of the upturn). We noted that, in the presence of 10 kOe, the peak intensity is reduced with a marginal shift (~ 0.2 K) of the peak towards low-$T$ range and this curve is plotted in the figure for the sake of clarity. This reduction of the peak temperature is consistent with AF ordering. On the other hand, for $H=$ 30 and 50 kOe, the peak shits towards higher $T$-range, which is supportive of a tendency towards ferromagnetic alignment. We have derived isothermal entropy change due to the attainment of these high fields (from zero field) and the $\Delta S$ values are plotted as an inset in Fig. 8c. While for $H=$ 10 kOe, the values are negligible at all temperatures, the values are quite significant in the temperature range 1.8 – 20 K for higher fields. The peak values are about 5 and 8.5 J/kg K for $H=$ 30 and 50 kOe respectively, marginally higher compared to that for the Ho compound.

In Figs. 9a and 9b, $\rho(T)$ and isothermal MR data are shown. The slope of the curve, $d\rho/dT$, remains positive in the entire $T$-range in the paramagnetic state (left inset of Fig. 9a), unlike in the Ho case. However, an application of magnetic field as small as 10 kOe depresses the values (however small it may be) in the range 3.5 – 15 K; for further higher values of $H$, the magnitude of the depression increases extending to much higher temperatures, as shown in Fig. 9a. Thus, in 50 and 70 kOe, distinct negative MR can be seen even at temperatures as high as 75 K, which is >20$T_N$. Thus, the magnitude of MR increases with decreasing temperature in the paramagnetic state. This naturally implies that, though there is no minimum in the raw data, the magnetic part (that is, spin-disorder contribution) appears to increase with decreasing temperature (see the right inset of Fig. 9a). [In the absence of a suitable reference for lattice contribution, this is a better way of inferring a qualitative picture about the temperature dependence of 4f contribution to $\rho$]. As one enters the magnetically ordered state, the spin-disorder contribution to $\rho$ tends to diminish, as a result of which there is a kink at $T_N$ in the zero-field curve; this kink in higher fields tends to get smeared and the change in slope gets shifted to a higher temperature - distinctly visible for $H=$ 50 and 70 kOe - due to the dominating tendency of ferromagnetic correlations. Isothermal MR data shown in Fig. 9b confirms that the sign of MR is negative in the entire temperature and field range of measurements. The absence of positive sign of MR (quadratically varying with $H$) at 1.8 and 3 K, expected for simple antiferromagnets establishes magnetic-gap effects in the Fermi surface. Monotonic increase of the magnitude of MR with increasing field or with decreasing temperature is also obvious from this figure. It is to be noted that there is a sudden increase in the slope of MR($H$) curve for 1.8 K around 15 kOe, suggesting the presence of spin-reorientation around this field, though it appears to be smeared in $M(H)$ data. MR is found to vary sluggishly at low fields at 1.8 and 3 K (almost quadratically) and the sharper drop expected [28] for spin-glasses and ferromagnets is absent; thus, this transport behavior is complex. Needless to emphasize that MR varies quadratically at 8, 15 and 20 K as expected for paramagnetism. But the magnitude of magnetoresistive response is quite large far above $T_N$, e.g., about -10% at 20 K for $H=$ 70 kOe, qualitatively similar to the case of Ho.

### 4. Discussions

From above, it is clear that the two compounds, Ho$_4$RhAl and Er$_4$RhAl, order magnetically at 8.8 and 3.5 – 4.0 K respectively. The spread of about 0.5 K for the latter is due to the fact that both AF and spin-glass transitions nearly seem to match. Depending on the sensitivity of measurements to the onset of different magnetic states, one gets signatures of transitions at slightly different temperatures. Gd$_4$RhAl and Tb$_4$RhAl [Ref. 12] order magnetically at ~46 and ~32 K with the latter obeying de Gennes scaling; however, the observed $T_N$ values for the title compounds are reduced with respect to the de Gennes scaled values (13.1 and 7.4 K). Usually, crystal-field effects and/or frustration effects have been known to contribute to this reduction.

We summarize below various other key features and also compare and contrast with the properties reported by us, in particular, for isostructural compounds.

- Ho$_4$RhAl exhibits an additional magnetic transition at about 5 K, which is of spin-glass type (that is, reentrant spin-glass behavior). This is similar to that observed for Gd$_4$RhAl ($T_N=$ ~46 K; $T_G=$ ~21 K), Tb$_4$RhAl (~32 K; ~28 K), Gd$_4$PtAl (~64; ~20 K). In the case of Tb$_4$PtAl and Dy$_4$PtAl [Ref. 11], more than one spin-glass transition below respective magnetic ordering onset temperatures was found. [Incidentally, this Dy compound alone in this family exhibits

- ferromagnetic order, rather than AF order, which is by itself interesting]. We report here that, for Er$_4$RhAl, we are not able to resolve additional magnetic transitions above 1.8 K; however, the spin-glass features are found to set in close to (or at) $T_N$, and similar finding was reported for Tb$_4$PtAl [Ref. 10]. Clearly, these compounds provide a variety with respect to magnetic order, possibly serving as a playground to explore magnetic phase-separation concept due to competing magnetism among the three magnetic sites.
- As in the case of other members of this stoichiometry, the reduced magnetic transition temperature value with respect to that of $\theta_p$ may have its main origin in magnetic frustration. The observed spin-glass features are proposed as the evidence for this interpretation. We tend to believe that the frustration resulting in glassy features is due to complex magnetic interaction among the three sites through RKKY interaction, possibly leading to frustrated Q-vector. In this sense, in our opinion, such a glassy-like state is subtly different from conventional spin-glasses arising from disorder. The idea of unconventional glassy state was also mooted in the recent past to explain the magnetic features for an intermetallic kagome lattice, Tb$_3$Ru$_4$Al$_{12}$, and also for a triangular spin-chain insulating system, Ca$_3$Co$_2$O$_6$, exhibiting complex spin structures and slow order-disorder transformation respectively [29]; the spin-glass-like features in these materials (observed in both polycrystals and single crystals) were found to be anisotropic and intrinsic to geometrical frustration, but not explainable in terms of disorder. Similarly, studies on high quality disorder-free single crystals of the present materials is desirable.
- It is important to stress that no glassy-like anomaly was observed in the bulk measurements for Gd$_2$PdSi$_3$, for example, in ac and dc $\chi$ [3-5]. Therefore, the present systems could be candidates for glassy anomalies as one kind of manifestation of the frustration, induced by itinerant electrons [1]. It is not out of place to mention that the compound Tb$_2$PdSi$_3$ within the 2-1-3 family has been shown to exhibit anisotropic re-entrant spin-glass-like anomalies in bulk measurements [30] with a complicated magnetic-structural features [31], which is worthwhile to explore further in the context of RKKY interaction frustration.
- In the field-range of measurements, isothermal $M$ data suggest the existence of at least one spin-reorientation for title compounds below 70 kOe. But, these are not disorder-broadened first-order transitions, as these $M(H)$ and $\rho(H)$ curves are not hysteretic, unlike the situation for many other compounds in which case the virgin curves also lie outside respective envelope curves across first-order transitions.
- We now focus on the electrical transport behavior in the paramagnetic state, relevant to the concept (ii) briefed in the Introduction. In all the Rh containing compounds of this family, the spin-disorder contribution seems to gradually increase with decreasing temperature till respective magnetic ordering temperature (rather than remaining constant), as inferred from the suppression of $\rho$ by the application of magnetic fields. In the case of Ho compound, an upturn can even be seen in the paramagnetic state (just above $T_N$) in the zero-field data, as a direct evidence for the same. As a result, negative MR persists over a wide $T$-range above $T_N$. Thus, these compounds belong to a special class of metallic heavy rare-earth compounds, e g., GdNi, Gd$_2$PdGe$_3$, R$_2$PdSi$_3$, R$_3$Ru$_4$Al$_{12}$, R$_2$RhSi$_3$ [32], R$_4$RhAl (Gd and Tb) [Ref. 12] and in particular Gd$_2$PdSi$_3$ [Ref. 3], Gd$_2$CuGe$_3$ [Ref. 23], Gd$_3$RuSn$_6$ [26], whose transport behavior is Kondo-like. *In the compounds like R$_7$Rh$_3$ [Ref. 24] and Tb$_5$Si$_3$ [Ref. 25], negative d$\rho$/dT and large (negative) MR extends even up to room temperature.* All these are possible examples for the recent theories [2] for the RKKY interaction-induced transport anomalies due to classical spin-liquid behavior [2] before long range magnetic order sets in. As a result of such an interplay, various nanoscale collective magnetic states possibly compete in a complex fashion [33]. There is also a recent theoretical approach in terms of cluster magnetism [34] leading to a classical spin-liquid phase in the paramagnetic state; these authors propose that clusters may grow in size with decreasing temperature, when intercluster interaction gradually dominates intracluster interaction. We have earlier proposed [6] to explore whether magnetic skyrmions form randomly already in the paramagnetic state, leading to above transport

- anomalies, just as Kondo-impurity to Kondo lattice transformation takes place in the Kondo-lattices. We now learnt [35] that such random skyrmions are quite commonly seen in the simulations based on Kondo lattice models for large local spins in the classical spin-liquid region. It may be noted that, in the case of $Dy_4PtAl$ and $Gd_4PtAl$ also, there is such a distinct $\rho(T)$ minimum (though not in $Tb_4PtAl$), but the sign of MR interestingly is positive above $T_N$; it is not clear to us at present, in these Pt materials, whether this signals gradual domination of AF component (rather than ferromagnetic component) or whether it is due to the domination of classical conduction electron contribution over spin-disorder part in these Pt-based materials in the presence of external $H$. It is even surprising to note that a $\rho(T)$ minimum appears when a magnetic field is applied for $Tb_4PtAl$ [10]. The purpose of listing the behavior of these compounds presenting various scenario is to bring out peculiar electrical transport anomalies in the paramagnetic state of some heavy rare-earth compounds with different crystal structures to the attention of the community.

- Transport anomalies in the magnetically ordered state as well are intriguing in this family. *In $Ho_4PtAl$, there is no drop of $\rho$ below $T_N$ down to 1.8 K.* In sharp contrast to this, in the Gd and Tb analogues [12], the $\rho$ at 1.8 K is reduced by about 15 to 25% with respect to that at respective $T_N$, clearly establishing dramatic reduction with respect to de Gennes scaled values. Such a suppression of $\rho$-drop at $T_N$ in metallic systems is usually attributed to magnetic-gaps in Fermi surface [27]. Clearly, this behavior resembles that of $Gd_2PdSi_3$ [Ref. 3] for which non-trivial Fermi surface topology was established by ARPES [7]. *This is the key comparative point* for the applicability of concepts involving Fermi surface below $T_N$ for the present Ho case. Even in the case of $Er_4RhAl$, the magnitude of the drop below $T_N$ is negligible (<2%) compared to that in Tb and Gd analogues. The fact that MR remains negative offers support for the fact the AF state is not a simple one, also viewing together with sluggish field-dependence of MR at low-fields. Magnetic state in other compounds in this family are equally complex, noting that there are sign-cross-overs [9-12] in isothermal MR, implying strong competition between various ground states, even for small applications of magnetic field or a variation of temperature. Clearly, the transport behavior of magnetically ordered states of these compounds also present a variety to explore itinerant electron induced frustration effects, viewed together with the fact that there are spin-glass anomalies. It is of interest to carry out further studies on single crystals, particularly Hall anomalies, to explore possible antiferromagnetic skrymion behavior in these compounds.

- Just as MR(*T*) curves, $\Delta S(T)$ curves also have a long tail extending over a large temperature range above $T_N$, as reported originally for $Gd_2PdSi_3$ [Ref. 36], and such a correlation has been known to exist even in the anisotropy in these properties, e.g., in $Tb_2PdSi_3$ [Ref. 37]. Following such initial reports [36-39], also in elaborated in Ref. 12, a mean-field model connecting both these quantities was also proposed for the (cubic) $RAl_2$ family [40]. For other reports on this aspect, the readers may see Refs. 41-44. We plan to address this issue further in a future publication.

- An important point we stress here is that the peak values of isothermal $\Delta S$ is large compared to that for Gd analogue, as in the case of the Tb compound [12]. These materials therefore offer a platform to explore a possible scenario proposed by us recently [12] that aspherical nature of 4f orbital might promote better MCE under favorable circumstances; that is, MCE could be controlled by single-ionic orbital effects.

## 5. Conclusions

We have presented the results of ac and dc magnetization, heat-capacity, electrical resistance and magnetoresistance measurements as a function of temperature down to 1.8 K for the two compounds, $Ho_4RhAl$ and $Er_4RhAl$, hitherto not paid much attention in the literature. We have compared and contrasted the observed properties with those of some other heavy rare-earth members in this family, pointing out a number of interesting features. In particular, the absence of a pronounced depression of the drop in $\rho$ at $T_N$ down to 1.8 K, similar to that in $Gd_2PdSi_3$ (distinctly in Ho compound), is a key observation, by analogy, speaking in favor of the role of itinerant interactions on the complexities in the magnetic ordering as in this Gd compound. We think that the observed glassy anomalies in magnetic measurements for the present cases is one manifestation of the magnetic frustration induced by itinerant electrons. It is worthwhile to carry out ARPES and band structure studies on the 4:1:1 ternary compounds as well as neutron diffraction studies to understand the magnetic structure as a function of temperature and magnetic field. We conclude that these heavy rare-earth members with strictly localized 4f electrons are rich in their magnetic and transport properties, serving as testing grounds for the recent theoretical concepts, bringing out newly realized roles of itinerant electrons interactions in the magnetically ordered states as well as in the paramagnetic state [1, 2]. In this context, we have also briefly reviewed the Kondo-like electrical resistivity and magnetoresistance anomalies in the paramagnetic state of some heavy rare-earth compounds in different crystallographic symmetries, reported by one of us during last few decades. Finally, we have also shown similarities in the magnetoresistance and magnetocaloric effect in the compounds under discussion. We hope these will motivate additional theoretical approaches.

**Acknowledgement**

The support of Kartik K Iyer while doing the experiments is gratefully acknowledged. The authors would like to thank Z. Wang, C.D. Batista and S. Hayami for their helpful comments. One of the authors (E.V.S) would like to thank Department of Atomic Energy, Government of India, for awarding Raja Ramanna Fellowship.

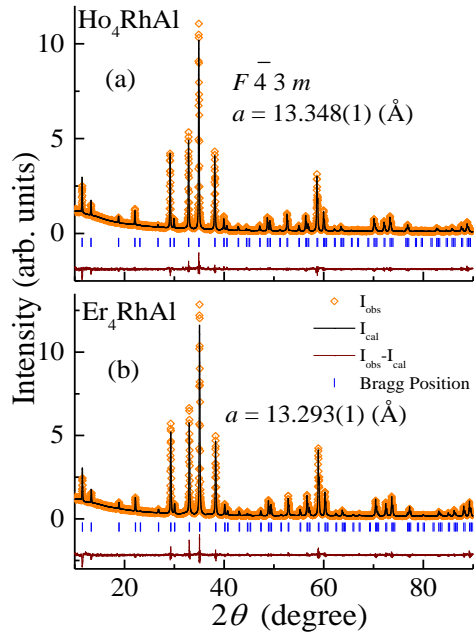

Fig. 1: X-ray diffraction patterns (Cu $K_\alpha$) of $Ho_4RhAl$ and $Er_4RhAl$ at 300 K. Rietveld fitting results (the continuous lines through the data points and fitted parameters) along with the positions expected for diffraction lines (vertical bars) and the difference between experimental and fitted line are shown.

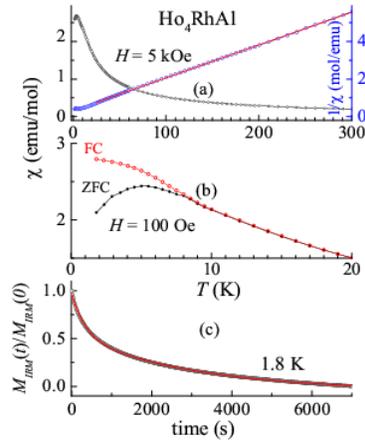

Fig. 2: Temperature dependent magnetic susceptibility ($\chi$) and inverse susceptibility obtained in a field of (a) 5 kOe by zero-field-cooling (ZFC) and (b) 100 Oe by ZFC as well as field-cooling (FC) for $Ho_4RhAl$. While the lines through the data points serve as guides in all curves, the line in inverse $\chi$ plot represents Curie-Weiss fitting. (c) Isothermal remnant magnetization as a function of time obtained for 1.8 K.

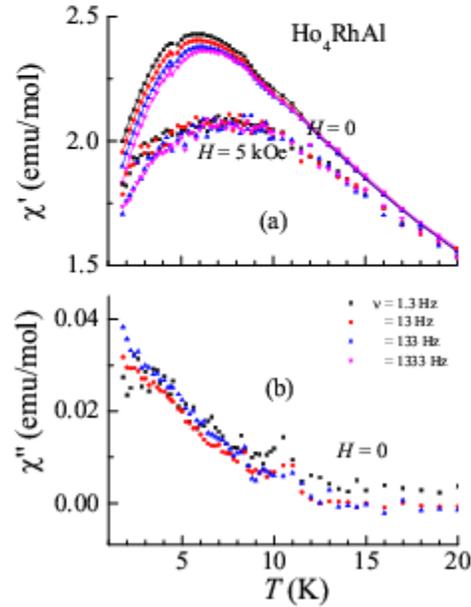

Fig. 3: Temperature dependence of real ($\chi'$) and imaginary ($\chi''$) parts of ac susceptibility as a function in the range 1.8-20 K for $Ho_4RhAl$ in zero field (and in 5 kOe for $\chi'$), measured with different frequencies (1.3, 13, 133 and 1333 Hz).

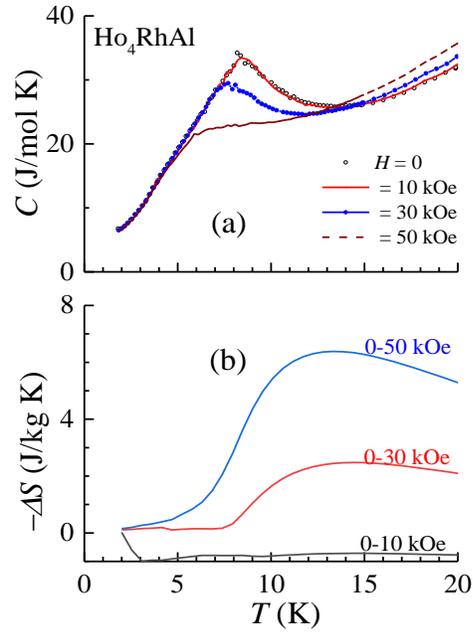

Fig. 4: Heat-capacity (*C*) as a function of temperature (1.8 – 20 K) for Ho$_4$RhAl in the absence of magnetic field as well as in 10, 30 and 50 kOe. The lines through the data points serve as guides to the eyes. (b) Isothermal entropy change brought out by changes in the magnetic field.

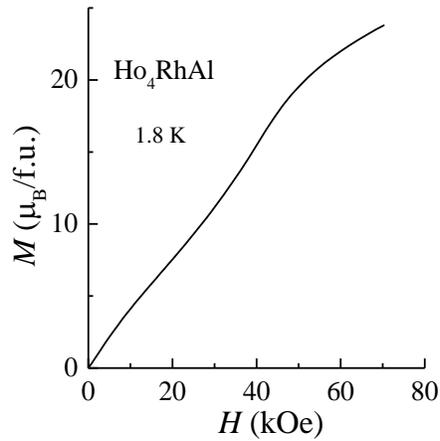

Fig. 5: Isothermal magnetization per formula unit (f.u.) (virgin) curves as a function of magnetic field at 1.8 K for Ho$_4$RhAl.

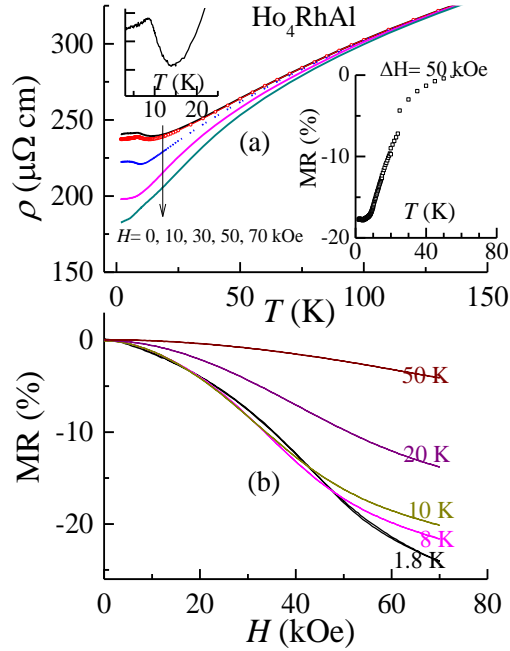

Fig. 6: Temperature dependence of electrical resistivity for Ho$_4$RhAl in the presence of different external magnetic fields. Left inset shows the zero-field profile at low temperatures to highlight the existence of a minimum. Right inset shows the magnetoresistance derived for 50 kOe. (b) Isothermal magnetoresistance behavior at 1.8, 8, 10, 20, and 50 K (up to 70kOe).

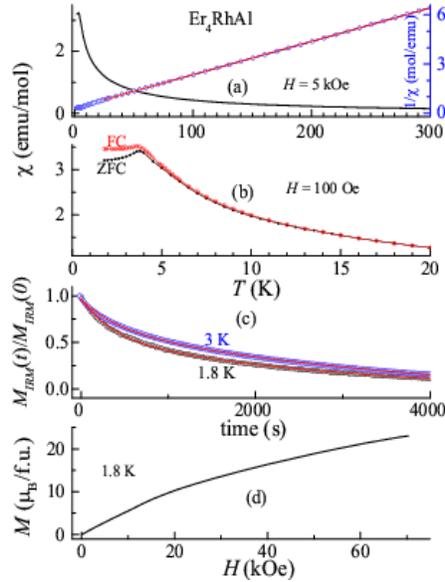

Fig. 7: For Er$_4$RhAl, magnetic susceptibility and inverse susceptibility obtained in a field of (a) 5 kOe by zero-field-cooling (ZFC) and (b) 100 Oe by ZFC as well as field-cooling (FC). The lines through the data points serve as guides in all curves in general, though in inverse $\chi$ plot, the line is obtained by Curie-Weiss fitting. (c) Isothermal remnant magnetization as a function of time for 1.8 and 3 K. (d) Isothermal magnetization per formula unit (f.u.) (virgin curve) at 1.8 K.

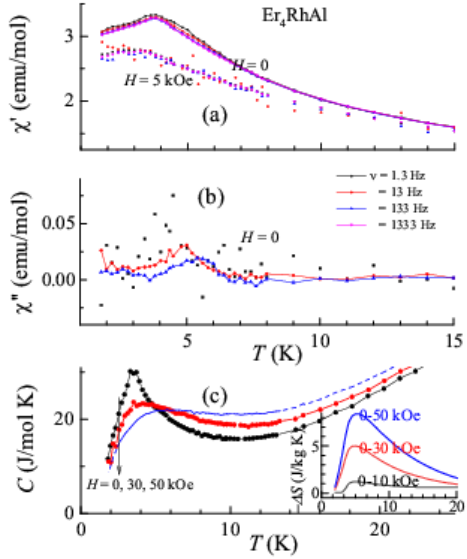

Fig. 8: Real (χ') and (b) imaginary (χ") parts of ac susceptibility as a function of temperature (1.8 - 15 K) for Er$_4$RhAl in zero field (and in 5 kOe for χ'), measured with different frequencies. χ" curves in 5 kOe are featureless and hence are not shown. (c) Heat-capacity behaviour in the range 1.8 – 30 K in 0, 30 and 50 kOe. Isothermal entropy change is plotted in the inset of Fig. 8c.

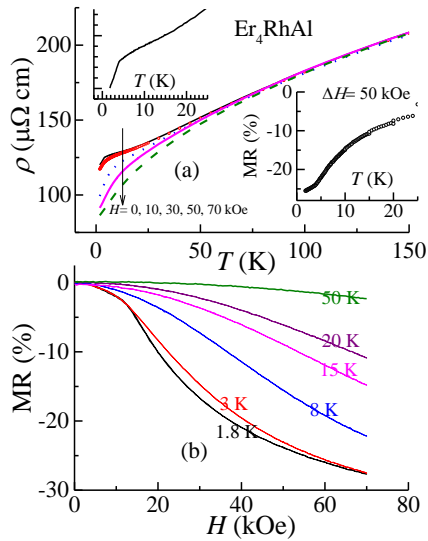

Fig. 9: Temperature dependence (1.8 - 300 K) of electrical resistivity for Er$_4$RhAl in the presence of different external magnetic fields. Left inset shows the zero-field data below 25 K to bring out the absence of a minimum. Right inset shows the magnetoresistance derived for 50 kOe. (b) Isothermal magnetoresistance behavior at 1.8, 3, 8, 15, 20, and 50 K (up to 70kOe).

15